\documentclass[default, iicol]{sn-jnl}
\usepackage{lineno}
\usepackage{subcaption}
\usepackage{csquotes}
\usepackage[mathscr]{eucal}
\usepackage{setspace}
\usepackage{multirow}
\usepackage{longtable}
\usepackage{graphicx}
\usepackage[ruled,vlined]{algorithm2e}
\usepackage{stfloats}

\usepackage{amsmath,amssymb,amsfonts}%
\usepackage{amsthm}%
\usepackage{mathrsfs}%
\usepackage[title]{appendix}%
\usepackage[dvipsnames]{xcolor}%
\usepackage{textcomp}%
\usepackage{manyfoot}%
\usepackage{booktabs}%
\usepackage{listings}%

\usepackage[nameinlink]{cleveref}

\newcommand{\red}[1]{#1}

\begin{document}

\title[Article Title]{Rapid Artefact Removal and H\&E-Stained Tissue Segmentation}

\author*[1,2]{B. A. Schreiber}\email{bas43@cam.ac.uk}
\author[1,2,3]{J. Denholm}
\author[1,3]{F. Jaeckle}
\author[4]{M. J. Arends}
\author[5]{K. M. Branson}
\author[2,3]{C.-B. Sch{\"o}nlieb}
\author*[1,3]{E. J. Soilleux}\email{ejs17@cam.ac.uk}

\affil*[1]{\orgdiv{Department of Pathology}, \orgname{University of Cambridge}, \orgaddress{\street{Tennis Court Road}, \city{Cambridge}, \postcode{CB2 1QP}, \state{Cambridgeshire}, \country{United Kingdom}}}
\affil[2]{\orgdiv{Department of Applied Mathematics and Theoretical Physics}, \orgname{University of Cambridge}, \orgaddress{\street{Wilberforce Road}, \city{Cambridge}, \postcode{CB3 0WA}, \state{Cambridgeshire}, \country{United Kingdom}}}
\affil[3]{\orgname{Lyzeum ltd}, \orgaddress{\city{Cambridge}, \postcode{CB1 2LA}, \state{Cambridgeshire}, \country{United Kingdom}}}
\affil[4]{\orgdiv{Edinburgh Pathology, Institute of Genetics \& Cancer}, \orgname{University of Edinburgh}, \orgaddress{\street{Crewe Road}, \city{Edinburgh}, \postcode{EH4 2XR}, \country{UK}}}
\affil[5]{\orgdiv{Artificial Intelligence and Machine Learning}, \orgname{GSK plc}, \orgaddress{\street{Great West Road}, \city{Brentford}, \postcode{TW8 9GS}, \state{Middlesex}, \country{UK}}}

\abstract{We present an innovative method for rapidly segmenting hematoxylin and eosin (H\&E)-stained tissue in whole-slide images (WSIs) that eliminates a wide range of undesirable artefacts such as pen marks and scanning artefacts. Our method involves taking a single-channel representation of a low-magnification RGB overview of the WSI in which the pixel values are bimodally distributed such that H\&E-stained tissue is easily distinguished from both background and a wide variety of artefacts. We demonstrate our method on 30 WSIs prepared from a wide range of institutions and WSI digital scanners, each containing substantial artefacts, and compare it to segmentations provided by Otsu thresholding and Histolab tissue segmentation and pen filtering tools. We found that our method segmented the tissue and fully removed all artefacts in 29 out of 30 WSIs, whereas Otsu thresholding failed to remove any artefacts, and the Histolab pen filtering tools only partially removed the pen marks. The beauty of our approach lies in its simplicity: manipulating RGB colour space and using Otsu thresholding allows for the segmentation of H\&E-stained tissue and the rapid removal of artefacts without the need for machine learning or parameter tuning.}


\maketitle
\section{Introduction}\label{sec:intro}
\begin{figure*}[ht!]
    \centering
    \begin{subfigure}[t]{.45\textwidth}
        \centering
        \frame{\includegraphics[width=\textwidth]{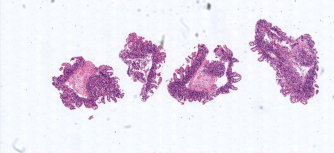}}
        \subcaption{}\label{fig:clean_biopsy}
    \end{subfigure}
    \hfill
    \begin{subfigure}[t]{.45\textwidth}
        \centering
        \frame{\includegraphics[width=\textwidth]{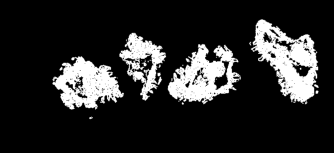}}
        \subcaption{}\label{fig:clean_mask}
    \end{subfigure}
    \hfill
    \begin{subfigure}[t]{.45\textwidth}
        \centering
        \frame{\includegraphics[width=\textwidth]{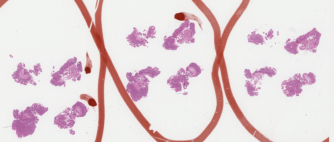}}
        \subcaption{}\label{fig:pen_mark_biopsy}
    \end{subfigure}
    \hfill
    \begin{subfigure}[t]{.45\textwidth}
        \centering
        \frame{\includegraphics[width=\textwidth]{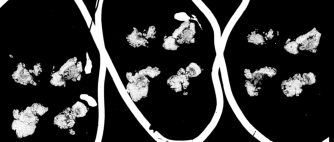}}
        \subcaption{}\label{fig:pen_mark_otsu}
    \end{subfigure}
    \caption{(a) A WSI of a H\&E-stained biopsy containing only minor, non-interfering artefacts.
        (b) The tissue segmentation provided by applying Otsu thresholding to the \red{luminance} of the WSI. The clear distinction between the intensities of the tissue and the rest of the WSI caused the Otsu threshold to lie between the maximum intensity of the tissue and the minimum intensity of background, allowing for a successful tissue segmentation.
        (c) A WSI of an H\&E-stained biopsy containing orange pen marks splitting the tissue of different levels and used to identify three features of interest.
        (d) The tissue segmentation provided by applying Otsu thresholding to the \red{luminance} of the WSI. The pen marks interfered with the Otsu threshold calculation, resulting in a tissue segmentation that contains tissue and pen marks.}\label{fig:otsu}
\end{figure*}
Otsu thresholding~\cite{otsu_threshold_1979} is often applied to the \red{luminance} of whole-slide images (WSI) of haematoxylin and eosin (H\&E)-stained tissue for the purposes of segmentation~\cite{wang_weakly_2020, denholm_multiple-instance-learning-based_2022, campanella_clinical-grade_2019, khened_generalized_2021,anghel_high-performance_2019,haghighat_automated_2022, smith_developing_2021,veta_predicting_2019, schmauch_transcriptomic_2019, zhang_dtfd-mil_2022} (see Figure~\ref{fig:otsu}), including in popular histopathological analysis tools Histolab~\cite{marcolini_histolab_2022} and PyHist~\cite{MunozAguirre2020}.
However, Otsu thresholding only successfully segments the tissue from the background when the tissue and background pixels are well-separated in a greyscale representation of the WSI.
While this is often the case in artefact-free WSIs, WSIs often contain artefacts such as pen marks and dark scanning artefacts, which cause this assumption to fail, thus resulting in artefacts wrongly identified as tissue, tissue rejected as background, or both (see Figure~\ref{fig:otsu}). While there are a large and diverse range of artefacts that can occur on a WSI, in the context of this paper artefacts will refer only to pen marks (see Figures~\ref{fig:overview}a, b, c, d, and g), bounding boxes added by the scanners (see Figures~\ref{fig:overview}e, f and g) scanning artefacts such as dark blobs or text (see Figures~\ref{fig:overview}e and f).

The exclusion of pen marks in particular is a crucial first step for any machine learning-based automated WSI analysis pipeline;
pathologists often use pen marks to highlight areas of interest which, if observed by a machine learning algorithm, could result in deleterious bias, spurious classifications or even data leakage, thus reducing the confidence in the performance metrics and the generalizability of the algorithm~\cite{kaufman_leakage_2012}.

In this paper, we propose a new tissue segmenting algorithm for H\&E-stained tissue which can segment tissue in the presence of artefacts. We tested our method on WSIs of H\&E-stained duodenal biopsies prepared at multiple different institutions, scanned using multiple different scanners, and containing a large range of artefacts of different types, shapes and colours.

\section{Method}
Our method improves on Otsu thresholding by selecting a representation of the WSI data that better separates H\&E-stained tissue from background and artefacts than \red{luminance}. Given a three channel image $I = [I_{R}, I_{B}, I_{G}]$, the channels are normalized so that the channels of each pixel are represented by floats ranging from 0 to 1. Then, the following representation of the data is calculated:

\begin{equation}\label{eq:he_otsu}
    T = \textrm{ReLU}(I_{R} - I_{G})\odot\textrm{ReLU}(I_{B} - I_{G})
\end{equation}
where $\textrm{ReLU}(x) = \max(x,0)$ is the rectifier linear unit
and $\odot$ is the Hadamard product, both of which act element-wise.
Otsu thresholding is then used to separate tissue and non-tissue pixels~\cite{otsu_threshold_1979}.
Note that this calculation requires no parameter training or tuning.
A Python implementation of this previously unreported algorithm can be found here~\url{https://gitlab.developers.cam.ac.uk/bas43/h_and_e_otsu_thresholding}
in accordance with the Guidelines for Authors Submitting Code and \& Software presented in Nature Research~\url{https://www.nature.com/nature-portfolio/editorial-policies/reporting-standard#reporting-requirements}.
All relevant guidelines were followed in the development and testing of this algorithm.

\begin{figure*}[ht!]
    \centering
    \includegraphics[width=\textwidth]{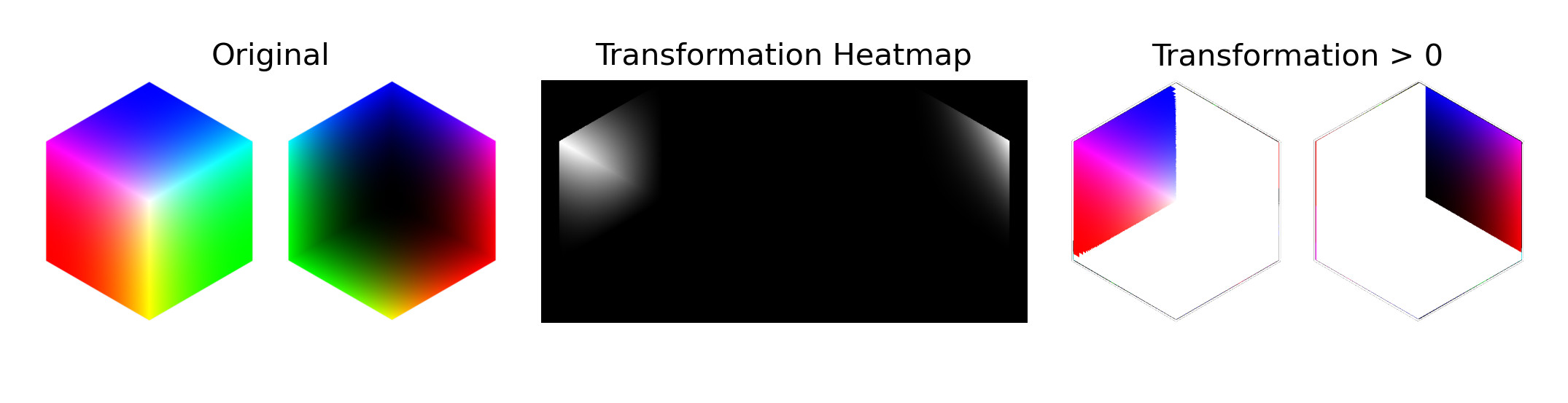}
    \caption{\red{Left: Two 24-bit colour cubes, one with the white corner at the origin and one with the black corner at the origin.
    Middle: The pixels which result in values greater than 0 in the representation specified in Equation~\ref{eq:he_otsu} and Algorithm~\ref{alg:he_otsu}.
    Right: The pixels that have values greater than 0 in this representation}}\label{fig:colour_cubes}
\end{figure*}

\begin{algorithm}[b!]
    \SetAlgoLined
    RGB Image: $[I_{R},I_{G},I_{B}]$ \\
    Normalize: $I_{R},I_{G},I_{B} \gets I_{R}/255,I_{G}/255,I_{B}/255$\\
    R - G Representation: $I_{R-G} \gets \textrm{ReLU}[I_{R} - I_{G}]$ \\
    B - G Representation: $I_{B-G} \gets \textrm{ReLU}[I_{B} - I_{G}]$ \\
    Tissue Representation: $T \gets I_{R-G} \odot I_{B-G}$ \\
    Otsu threshold: $\gamma \gets \mathrm{Otsu}[T]$ \\
    \eIf{$T[p] > \gamma$}{Pixel $p$ is segmented as tissue}{Pixel $p$ is rejected}
    \caption{Our method for segmenting H\&E stained tissue}\label{alg:he_otsu}
\end{algorithm}
The assumption made by Otsu thresholding is that tissue and non-tissue pixels can be separated by their grey-scale values, which is not the case when artefacts are present. However, our method, which is described in Equation~\ref{eq:he_otsu} and Algorithm~\ref{alg:he_otsu}, is based on the assumption that the tissue pixels can be identified by being \textit{both} more blue than green and more red than green as compared to non-tissue pixels. The advantage of our method is that all shades of grey have approximately the same value in the red channel as the green channel, so their difference is 0, while pixels of H\&E-stained tissue have higher values in the blue and red channels than the green. Setting all negative values in both representations to zero ensures that artefacts with high green channels compared to blue or red channels do not adversely influence the threshold calculation, and are thus considered as background. Thus, this representation results in a bimodal distribution that separates pixels that are the most ``purple-pink'' from others, so pen marks (which are often black, blue, green or red) are also excluded, independent of the pixel's light intensity. \red{Pixels on an RGB colour cube that have a non-zero value in this representation are shown in figure~\ref{fig:colour_cubes} and comparisons between Otsu thresholding and our method on an RGB colour cube can be seen in the supplementary material.}

We compared the performance of our method against Otsu thresholding and Histolab's pen filtering tools by applying these methods to a dataset of WSIs and assessing the resulting tissue segmentations qualitatively.

\subsection{Data}
To compare the performances of the Otsu thresholding, Histolab and our method, we applied both methods to a selection of 60 WSIs of H\&E stained duodenal biopsies.
Of the 60 WSIs selected:
\begin{itemize}
    \item 15 contained pen marks
    \item 15 contained scanning artefacts
    \item 30 contained no significant artefacts
\end{itemize}

The WSIs were hand-picked so that they contained a wide range of artefacts of different types, shapes and colours. The WSIs were scanned with a wide range of digital scanners (Ventana, Aperio, Hamamatsu and Philips), and the 30 WSIs with no significant artefacts were selected at random and matched for scanner type of the 30 WSIs with pen marks or artefact.
 
\subsection{Ethical Statement}
All fully anonymized slide scans (and patient data) were obtained with full ethical approval from the Oxfordshire Research Ethics Committee A (IRAS\@: 162057; PI\@: Prof. E. Soilleux), and the method was performed in accordance with their guidelines and regulations. Informed consent was obtained from all subjects and/or their legal guardian(s).

\section{Results}

\begin{figure*}[hp!]
    \centering
    \includegraphics[width=0.9\linewidth]{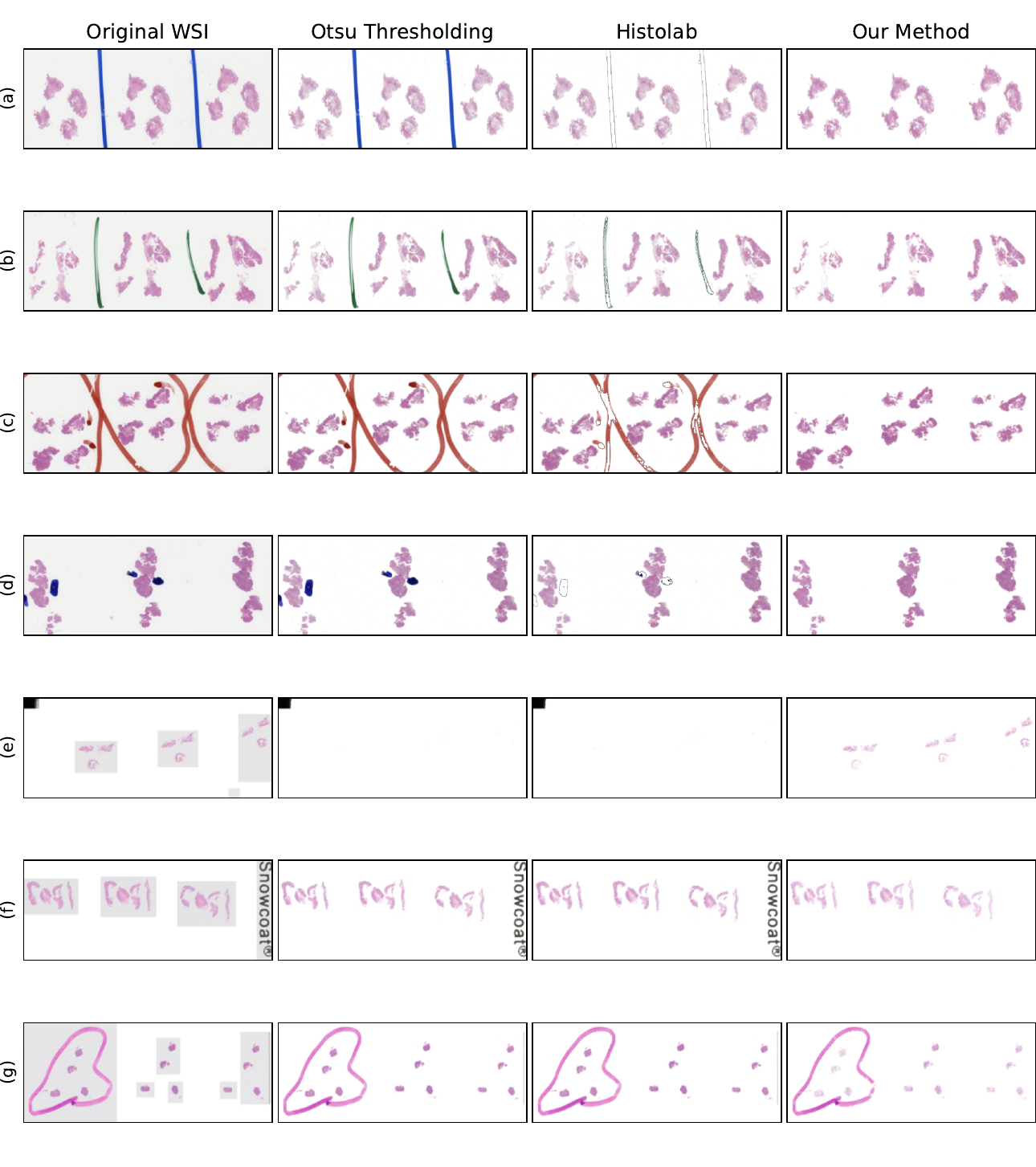}
    \caption{Seven WSIs of H\&E-stained biopsies containing artefacts of a wide range of types and colours.
        The aim was to segment the tissue without including background and artefacts.
        First Column: The original WSI.
        Second Column: The tissue segmentation provided by applying Otsu thresholding to the \red{luminance} of the WSIs placed on a white background. Otsu thresholding failed to reject a single artefact and failed to segment the tissue in (e).
        Third Column: The tissue segmentation from Histolab tissue thresholding and pen filters. While there was partial pen mark removal in (a), (b), (c) and (d), the pen marks were not fully removed in any image, no tissue was segmented in (e), no pen marks were removed in (g). 
        Forth Column: The tissue segmentation from our method placed on a white background.
        Our method successfully segmented all tissue and rejected all background and artefacts except the pen marks in (g). Our method failed to reject the pen marks in (g) because the pen is the same colour as the eosin.}\label{fig:overview}
\end{figure*}

    Otsu thresholding, Histolab and our method were used to segment the tissue from the 60 WSIs described above. Examples of the WSIs selected and the tissue segmentation of these methods can be viewed in Figure~\ref{fig:overview}. Examples of the tissue segmentation masks provided by Otsu thresholding and our method, and the S{\o}rensen–Dice coefficient's between the segmentations \red{and a manually segmented tissue mask} are displayed in the supplementary material. The tissue segmentations were assessed by a single observer, and considered ``successful'' if all the following were true:
\begin{itemize}
    \item All tissue was segmented
    \item All background was rejected from the segmentation
    \item All bounding boxes were rejected from the segmentation
    \item All artefacts were rejected from the segmentation
\end{itemize}

Otsu thresholding rejected pen and scanning artefacts from the tissue segmentation in 0/30 WSIs containing artefacts. In 2/30 WSIs containing artefacts, the influence the artefacts had on the threshold was so great that the tissue was not segmented as tissue (see Figure~\ref{fig:overview}e).

The Histolab pen filtering tool only partially removed pen marks in Figures~\ref{fig:overview}a-d, and removed no pen marks in Figure~\ref{fig:overview}g. Other artefacts such as scanning artefacts were not effected by the Histolab tools.

Our method segmented the tissue in all 60/60 WSIs and rejected all artefacts in 29/30 WSIs containing pen and scanning artefacts. The only WSI where pen marks were included in the tissue segmentation can be seen in Figure~\ref{fig:overview}g. Here our method failed to reject the pen marks because tissue and non-tissue pixels could not be separated through their “pinkness”, when the pen marks were also pink.

\section{Discussion}

While Otsu thresholding segmented the tissue in all artefact-free WSIs and most WSIs with artefacts, it identified all artefacts as tissue as well. In 2 out of 30 WSIs with artefacts, the presence of artefacts caused the threshold to ignore tissue and include background in the tissue segmentation as seen in Figure~\ref{fig:overview}e.

The Histolab pen filtering tools were applied to the tissue segmentations in order to remove the remaining pen marks. The filtering tools performed best on blue pen marks, as seen in Figures~\ref{fig:overview}a and d. However, the tools did not remove the edges of pen marks of all colours, and dailed to detect the majority of all green and orange pen marks, as seen in Figures~\ref{fig:overview}b and c respectively. The pink pen marks presented in Figure~\ref{fig:overview}g remained untouched. Additionally, the Histolab tools were not designed to remove scanning artefacts and bounding boxes so these features remained.

Our method, on the other hand, segmented the tissue in all WSIs and rejected artefacts in all WSIs containing artefacts but one. The only exception can be seen in Figure~\ref{fig:overview}g, which contained pink pen marks that caused all methods to fail.

The thresholding algorithm presented here is a rapid, reliable and easily implementable tissue segmentation and artefact removal tool for WSIs of H\&E-stained tissue. In machine learning tasks especially, this tool can be used as a preprocessing step that ensures artefacts do not cause the machine learning algorithm to train on irrelevant patches or patches that contain data leaking pen marks.

It should be noted that this method is built to segment H\&E-stained tissue only, and will not perform as intended on tissue which has been stained with stains that do not appear pink/purple. However, this method should be relatively simple to generalize to other stains by using representations of the WSI data that uniquely differentiate the stained tissue from background and artefacts, and will be studied in future research.

\section*{Funding}
This work was supported by
the Pathological Society [PKAG/924]
and GlaxoSmithKline [LEAG/781]

\section*{Author Contributions}

B. A. Schreiber devised the thresholding algorithm and wrote the manuscript. J. Denholm and F. Jaeckle independently tested and compared Otsu thresholding, Histolab, and the thresholding algorithm presented here. Histological expertise was provided by M. J. Arends and E. J. Soilleux. The project was initialized by E. J. Soilleux and supervised by K. M. Branson, C.-B. Sch{\"o}nlieb and E. J. Soilleux. All authors were given the opportunity to review and comment on the manuscript.

\section*{Data Availability}
The datasets of WSIs analysed during this current study have not publicly available due to the large size of the WSIs and legal considerations. However, low-level representations of the WSIs used in the study have been made available at~\url{https://gitlab.developers.cam.ac.uk/bas43/h_and_e_otsu_thresholding}.

\newpage

\bibliographystyle{elsarticle-num-names}
\bibliography{references}

\end{document}